\documentstyle[11pt]{article}
\newcommand{\mup}{multiplicity}   
\newcommand{\dis}{distribution}
\newcommand{\hp}{\hspace{5mm}}
\newcommand{\hs}{\hspace{1cm}}
\newcommand{\om}{\omega}
\newcommand{\al}{\alpha}

\newcommand{\ga}{\gamma}
\newcommand{\Ga}{\Gamma}
\newcommand{\beq}{\begin{equation}}
\newcommand{\eeq}{\end{equation}}
\newcommand{\bqa}{\begin{eqnarray}}
\newcommand{\eqa}{\end{eqnarray}}
\newcommand{\bqq}{\begin{eqnarray*}}
\newcommand{\eqq}{\end{eqnarray*}}
\newcommand{\bzs}{\left[ }
\newcommand{\ezs}{\right] }
\def\Journal#1#2#3#4{{#1} {\bf #2}, #3 (#4)}

\def\NCA{{\em Nuovo Cimento} A}
\def\NIM{\em Nucl. Instrum. Methods}

\def\NPB{{\em Nucl. Phys.} B}
\def\PLB{{\em Phys. Lett.}  B}

\def\PR{{\em Phys. Rev.} }
\def\PRD{{\em Phys. Rev.} D}

\begin{document}           
\title{KNO Scaling in the Neutral Pion Multiplicity Distributions for 
$\pi^- p$ interactions at 40 and 250 GeV/c.}
\author{ L. DI\'OSI, S. KRASZNOVSZKY and I. WAGNER\\\\
KFKI Research Institute for Particle and Nuclear Physics,\\
Hungarian Academy of Sciences,\\ 
POB 49, H-1525, Hungary}
\date{20 May, 1999}
\maketitle
\abstract{We analysed the binomial \mup\  moments of the 
neutral pions, using an extension of the generating functional technique
for detection losses. We applied this model-independent method to the
individual $\gamma$ weights of 10000 events of $\pi^- p$ interactions at 40 
GeV/c. We compared the obtained results to those of 250 GeV/c. We used the
FRITIOF and a shifted KW \dis \ to describe the data.}

\section*{Introduction}
Investigation of \mup\ \dis s have so far been done mostly for charged 
particles. A comprehensive study of functional forms and fits for data have
been lately reviewed in Warsaw~\protect{\cite{W}}. Less information is
available on production of $\pi^0$ meson. The $\pi^0$ decay product, gammas,
may be observed in bubble chamber with low efficiency. 

We shall analize the moments of \mup\ \dis s of $\gamma$-s for $\pi^- p$ and
$\pi^- n$ interactions at 40 GeV. We will use the data from the Dubna 2m
propane bubble chamber. The statistics includes about 10000 events for
$\pi^- p$ and 3600 events for $\pi^-n$ interactions. We have 25$\%$ mean
efficiency~\protect{\cite{B}}. Every individual conversion weight of $\ga$ is
at our disposal from the data summary tape~(DST).

Since the detection probability is lower than 100$\%$ the measured \dis \ is
different from the true one. The problem is the following: how to take into
account this difference in the analysis of the data.

\section{General method of the correction of detector losses.}
The generating functional technique is both extremely general and useful. 
On the one hand it can be used to prove important 
theorems~\protect{\cite{Br}}, on the other hand it permits the description of 
detection losses too~\protect{\cite{DL}}.
In order to give some insight into this problem we will show a general model
independent method by Di\'osi~\protect{\cite{DL}}. We shall recall some 
statements from these papers.

The true n-particle exclusive \dis \ 
$s^{(n)}$ with the proper normalization is the following:
\beq \int s^{(n)}\left( k_1 ,\ldots ,k_n\right) dk_1 ,\ldots dk_n =n!p_n
 \label{st}\eeq
where $p_n$ is the probability of fixed n \mup \ and $k_n$ is the momentum of
the n-th particle.

With the aid of the detection probabilities 
$\om =\om (k)$-s we can describe the measured exclusive \dis \
\bqa 
\bar{s}^{(n)}\left( k_1 ,\ldots ,k_n\right) &=& \sum_{m>n}\frac{1}{(m-n)!}
\int s^{(m)}\left( k_1 ,\ldots ,k_m\right)\om (k_1)\ldots \om (k_n)\cdot
\nonumber
\\ & & \cdot \tilde{\om} (k_{n+1})dk_{n+1} \ldots \tilde{\om} (k_m)dk_m
\eqa
where $\tilde{\om} = 1-\om$.

By definition the generating functional:
\beq F\bzs h(.) \ezs= \sum_{n=0}\frac{1}{n!}\int s^{(n)}\left( k_1 ,\ldots ,
k_n\right) h(k_1)dk_1\ldots h(k_n)dk_n \eeq
The exclusive \dis \ can be expressed by the derivatives of the generating
functional:
\beq s^{(n)}\left( k_1 ,\ldots ,k_n\right) = \left. \frac{\delta^{n}F}
{\delta h(k_1)\ldots \delta h(k_n)}\right| \begin{array}{c} {}\\h=0 
\end{array} \eeq
The generating functional of the measurable \dis :
\beq \bar{F}\bzs \bar{h}(.) \ezs= \sum_{n=0}\frac{1}{n!}\int \bar{s}^{(n)}
\left( k_1 ,\ldots ,k_n\right) \bar{h}(k_1)dk_1\ldots \bar{h}(k_n)dk_n \eeq
Using these eqs. the fundamental reconstruction formula can be obtained as:
\beq \bar{F}\bzs \bar{h}(.)\ezs =F\bzs \om (.)\bar{h}(.) + 1-\om (.)\ezs
\label{F1}\eeq
\beq F\bzs h(.)\ezs =\bar{F}\bzs \om^{-1}(.)h(.) + 1-\om^{-1} (.)\ezs
\label{F2}\eeq
We can check these formulae: if the argument $h(.)$ or $\bar{h}(.)=1$ then 
$\bar{F}[1]=F[1]$ for arbitrary $\om (.)$, and if $\om (.)=1$ then 
$F=\bar{F}$. We should note that (\ref{F1}) and (\ref{F2}) is a 
generalization of Nifenecker's results for the generating 
function~\protect{\cite{Ni}} if 
$h(.)\rightarrow z; \hs \om (.)\rightarrow$ constant, where the constant 
$\om$ is the neutron detector efficiency.

Armed with this techique we invert eq. (2):
\bqa s^{(n)}\left( k_1 ,\ldots ,k_n\right)&=&w(k_1)\ldots w(k_n)\sum_{i=0}
\frac{(-1)^i}{i!}\int \bar{s}^{(n+i)}\left( k_1 ,\ldots ,k_{n+i}\right)\cdot
\nonumber\\
& &\cdot \tilde{w}(k_{n+1})dk_{n+1}\ldots\tilde{w}(k_{n+i})dk_{n+i}
\eqa 
Where: \[w=\frac{1}{\om};\hs \tilde{w}=w-1\].
Taking a simple case, if $\om =$ constant then from eq. (2)
\beq \bar{p}_{\bar{n}}=\sum_{n\geq\bar{n}}p_n\left( 
\begin{array} {c} n\\ \bar{n} \end{array}\right) 
\om^{\bar{n}}(1-\om )^{n-\bar{n}}\label{PM} \eeq
where $\bar n$ is the measured \mup \ , and with
\[\tilde{w}=\frac{1}{\om}-1=-\left( 1-\frac{1}{\om}\right)\]
we can calculate from (8)
\[p_n =\frac{1}{n!}\sum_{i=0}\left( \frac{1}{\om}\right) ^n \left( 1-\frac{1}
{\om}\right) ^i \frac{1}{i!} \left( n+i\right) !\bar{p}_{n+i}\]
Substituting $n+i \rightarrow \bar{n}$ we obtain the so called Diven  
formula~\protect{\cite{Di}} for
\beq p_n=\sum_{\bar{n}\geq n}\bar{p}_{\bar{n}}\left( 
\begin{array}{c}  \bar{n}\\ n \end{array}\right) \left( \frac{1}{\om}
\right)  ^n \left( 1-\frac{1}{\om} \right) ^{\bar{n}-n}\eeq
If $\om$ is small we can arrive at a solution containing large oscillating and
sometimes even negative components of $p_n$\protect{\cite{Da}}. On the other
hand it was demonstrated that the moments ( $<n>$ and $D^2_n$ ) of the same
$p_n$ prove to be very stable in the case of multiplicities of fission 
neutrons~\protect{\cite{Da}}. We show these results in Table~1.
\begin{table}[ht]
\caption{True moments of \dis s for different 
experiments~\protect{\cite{Da}} }  
\vspace{0.4cm}
\begin{center}
\begin{tabular}{|c|c|r|c|c|}
\hline
No. exp.& $\om \%$ & No. events& $<n>$& $D^{2}_{n}$\\ \hline
$1$ & $48.3$ &  $7169$& $2.690\pm 0.036$& $1.388\pm 0.076$\\
$2$ & $48.2$ & $65015$& $2.690\pm 0.015$& $1.290\pm 0.025$\\ 
$3$ & $44.4$ &  $6928$& $2.690\pm 0.038$& $1.212\pm 0.084$\\
$4$ & $39.9$ & $20359$& $2.690\pm 0.025$& $1.173\pm 0.057$\\ 
$5$ & $23.7$ &  $4039$& $2.690\pm 0.071$& $1.230\pm 0.272$\\
$6$ & $22.0$ &  $4039$& $2.690\pm 0.075$& $1.587\pm 0.311$\\ \hline
\end{tabular}
\end{center}
\end{table}

The same conclusion can be drawn from analitical calculation for Poisson
\dis ~\protect{\cite{DL}}.

In the general case $\om =\om (k)$ we can 
prove~\protect{\cite{DL}} that the true binomial moment 
\beq B_j =\frac{1}{j!}\int w_1 \cdots w_j \bar{f}_j (w_1 ,\ldots ,w_j )dw_1
\cdots dw_j  \eeq
where $\bar{f}_j$ is the measured inclusive \dis .

\section{Gamma moments from the data summary tape}
$\om$ is the probability of $e^+ e^-$ pair creation of a secondary $\gamma$:
\beq \om =1-\exp \left( \frac{L_x}{L}\right) =\frac{1}{w} \eeq
where $L_{x}$ is the potential length, $L=L(k)$ is the radiation length and
$w$ denotes the conversion weight.
The general prescription for the true binomial moments~\protect{\cite{DL}} 
\beq B_k =\left< \tilde{B}_k \right> _{DST} \eeq
where $\tilde{B}$ is the following for every event:
\beq \tilde{B_k}=\left\{ \begin{array}{l} 0\hs if\hs \bar{n} <k\\
\sum_{\al}w_{i_1}\cdot w_{i_2}\cdots w_{i_k} \end{array} \right. \eeq 
where
\(\al= \left( \begin{array} {c} \bar{n}\\ k\end{array}\right)  \)
and $\bar{n}$ is the detected number of gammas in an event and the summation
goes for all the $\al $ different set of indices. 
E.g. $\bar{n}=4$,\\ $k=2,\hp \al =6$
\[\tilde{B}_2 =w_1 w_2 + w_1 w_3 + w_1 w_4 + w_2 w_3 + w_2 w_4 + w_3 w_4  \]
In addition to $B_k$ we calculated the errors and the correlations of $B_k$
from the DST:
\beq (\Delta B_k)^2 =\left< \left( \tilde{B}_k -B_k \right) ^2 \right> 
_{DST} \eeq
\beq \Delta B_k\Delta B_l =\left< (\tilde{B}_k -B_k)(\tilde{B}_l -B_l)\right> 
_{DST} \eeq 
Thus we obtained significant results for the first three binomial moments.
Assuming that all gammas come from neutral pions:
\[p_{2n}^{(\ga )}=p_n ^{(\pi^0)}, \hs p_{2n+1}^{(\ga )}=0 \]
we can calculate arbitrary $\pi^0$ moments using the proper generating 
function:
\[ G^{(\ga )}(z)=\sum p_n ^{(\ga )}z^n \] and 
\[ G^{(\pi ^0 )}(z)=\sum p_n ^{(\pi ^0 )}z^n = G^{(\ga )}(z^{\frac{1}{2}}) \]
In this way we calculated the average \mup\ in accordance with earlier 
published data~\protect{\cite{B}} and \(\frac{<n^{\pi^+}>+<n^{\pi^-}>}{2}=
\frac{2.18 + 2.81}{2} = 2.5\) which is equal to $<n^{(\pi^0)}>=2.49\pm 0.04$
for $\pi^- p$ interactions at 40GeV.
The behaviour of $\pi^- n$ data on $c_2 =1.69\pm 0.11$ and 
$c_3 =3.61 \pm 0.43$ are similar to $c_2 =1.64 \pm 0.07$ and 
$c_3 = 3.38 \pm 0.32$ found for $\pi^- p$ data.

We can compare our results 
with the 5m hydrogen bubble chamber data on $\pi^- p$ 
at 250 GeV~\protect{\cite{Dia}}. The statistics is
larger (20000 events) but $<\om>=14\%$ is smaller, the experimental detailes
have been described in~\protect{\cite{Dia}}.

At 250 GeV $c_2 =1.55\pm 0.12$ 
and $c_3 =3.04\pm 0.51$.

Within the errors the KNO moments $c_2$ and $c_3$
do not show a violation of KNO scaling~\protect{\cite{KNO}} between 40 and 250 
GeV.

\section{FRITIOF and shifted KW \dis \ for $\psi^{(\pi^0 )}$.}
We have generated 14500 FRITIOF events at every sample. The FRITIOF 
reproduces the mean multiplicities, the second scaled moments $c_2$
and $c_3$, which can be seen in Table 2. 
\begin{table}[ht]
\caption{Calculated parameters of shifted KW and FRITIOF \dis s for $c_3$}  
\begin{center}
\begin{tabular}{|r|c|c|c|c|c|c|}
\hline
& & & & & & \\
Experiment& $1+\epsilon$ & A& $<n>_{*}$& $c_3[pred]$& $c_3[exp]$& 
$c_3[FRITI] $\\ \hline
$\pi^- n\ \ 40\ GeV$ & $0.955$ & $0.71$ & $3.16$ & $3.57$ & $3.61\pm 0.43$ 
& $2.91\pm 0.11$\\
$\pi^- p\ \ 40\ GeV$ & $0.965$ & $0.75$ & $3.37$ & $3.38$ & $3.38\pm 0.32$ 
& $2.85\pm 0.10$\\ 
$\pi^- p\ 250\ GeV$ & $0.978$ & $0.77$ & $4.43$ & $2.98$ & $3.04\pm 0.51$ 
& $2.53\pm 0.09$\\ \hline
\end{tabular}
\end{center}
\end{table}

We should note that a shifted KW \dis \ has successfuly 
described~\protect{\cite{KW2}} 
the KNO moments and the \dis s for the single hemisphare data of DELPHI and 
OPAL collaborations. In order to predict the third scaled moments $c_3$--s we 
use a shifted KW \dis \ with parameter m=2, which has proved to be successful 
for charged particles in inelestic pp collisions~\protect{\cite{KW1,KW3}}. 
We carry out a shift with +1, since the KW \dis \ is equal to zero in n=0, 
and we use the so called stick approximation. It means that we use the 
continuous (denoted by *) KW
\dis 
\[ P_n ^* =\frac{m}{<n>_* \Ga (A)}F^A z^{ma-1}\exp\bzs -Fz^m \ezs \]
where
\[ F=\frac{\Ga ^m (A+\frac{1}{m})}{\Ga ^m (A)}\hs z=\frac{n}{<n>_*}\hs m=2\]
Taking the sum of $P_n^*$ for n=0,1,2,... we can define the remaining
$\epsilon$ in the Euler-MacLaurin formula:
\[ \sum_n P_n ^* (A,<n>_* )=1+\epsilon \] 
We can form the discrete pobabilities
\[P_n =\frac{P_{n+1}^* (A,<n>_* )}{1+\epsilon } \]
fulfilling the requirements: 
\[ <n>=\sum nP_n, \hs c_2 =\frac{\sum n^2 P_n}{<n>^2}\]
We display the calculated parameters $\epsilon$, $<n>_*$ and $A$ in Table~2.

Using these parameters we can predict $c_3$, which are in good agreement with 
the true expermental data.
With the shifted $P_{n+1}^*$ we create the continuous KNO function
\[ \psi(\tilde{z})=<n>_* P_{n+1}^* =(<n>+1)(1+\epsilon)^2 P_n (\tilde{z},A)\]
where
\[ <n>_* = (<n>+1)(1+\epsilon ),\hs \frac{n+1}{<n>_*} =\hat{z}\]
A KNO function with parameters A=0.743 and m=2 represents the calculated 
points In Fig.1. KNO scaling is seen as a function of $\hat{z}$ at two 
energies.

\section*{Conclusions}
\begin{enumerate}
\item The use of generating functional technique provides an elegant and
      concise derivation of formulae relating the true \dis \ function
      to the measured ones. With the aid of this general method we have
      immediately obtained Nifeneckers~\protect{\cite{Ni}} results on the 
      generator functions and Diven~\protect{\cite{Di}} formula for the true 
      \dis \ of fission neutron \mup , as a special case: $\om =$ constant.
\item Since the mean detection efficiency for $\ga $ is $25\%$ in
      propane (at 40 GeV) the reconstruction of the true \mup \ \dis \
      from the measured one is not efficient, but still the first three
      binomial moments of the true \dis \ can be obtained.
\item We have compared our results at 40 GeV with the published results at
      250 GeV. It was found that the KNO moments $c_2$ and $c_3$ of $\pi^0$
      are consistent with KNO scaling within their large (15 percent)
      errors up to third moment, in a model-independent way.
\item The experimental KNO moments are in agreement with the FRITIOF 
      simulation and with a shifted KW \dis \ containing the first two 
      binomial moments as an input. It is remarkable that this KW \dis \ 
      can predict the measured $c_3$. All the points of 
      $\pi^0$ \mup \ \dis s calculated
      as a shifted KW show a clear scaling curve between 40 and 250 GeV.
\end{enumerate}

\section*{Acknowledgment}
This work was supported by the Hungarian Academy of Sciences (Grant No. 
OTKA T 016377).

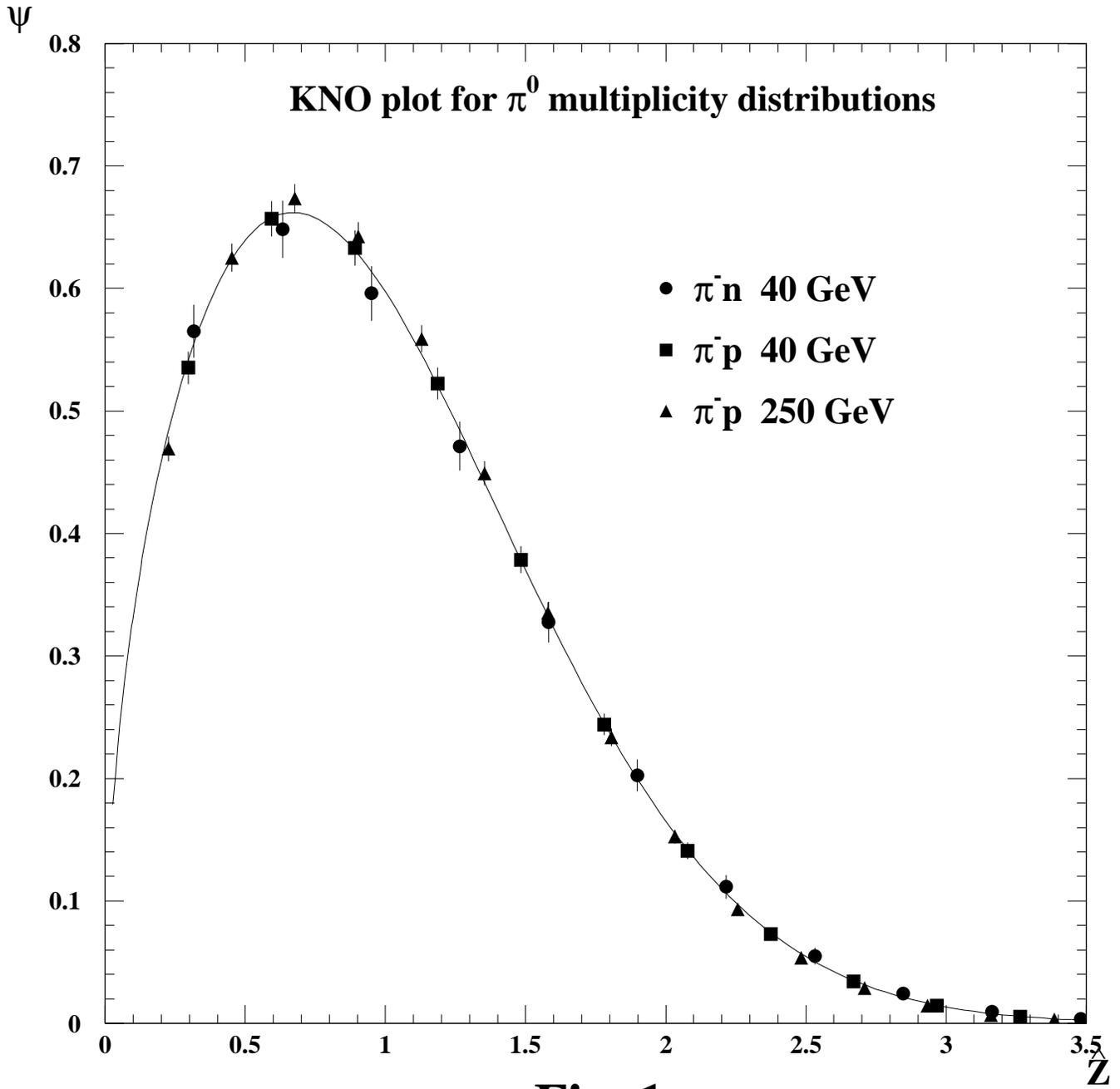
\end{document}